\documentclass[conference, a4paper]{IEEEtran}

\usepackage{amsmath,amssymb,amsfonts}
\usepackage{graphicx}
\usepackage{xcolor,cite}
\usepackage{booktabs}
\usepackage{nicefrac} 

\def\mathlette#1#2{{\mathchoice{\mbox{#1$\displaystyle #2$}}%
                           {\mbox{#1$\textstyle #2$}}%
                           {\mbox{#1$\scriptstyle #2$}}%
                           {\mbox{#1$\scriptscriptstyle #2$}}}} 
\renewcommand{\Vec}[1]{\mathlette{\boldmath}{#1}}  

\newcommand{\be}{\begin{equation}} 
\newcommand{\ee}{\end{equation}}
\newcommand{\ba}{\begin{array}}
\newcommand{\ea}{\end{array}}
\newcommand{\bdm}{\begin{displaymath}}
\newcommand{\edm}{\end{displaymath}}
\newcommand{\bea}{\begin{eqnarray}}
\newcommand{\eea}{\end{eqnarray}}
\newcommand{\bean}{\begin{eqnarray*}} 
\newcommand{\eean}{\end{eqnarray*}}
\newcommand{\me}{\text{e}}

\newcommand{\mj}{\text{j}}

\def\argmin{\mathop{\text{argmin}}}

\def\nTS{\ensuremath{T_\text{S}}} 

\def\oH{\ensuremath{^\text{H}}} 
\def\oT{\ensuremath{^\text{T}}} 

\def\npBS{\ensuremath{\Vec{p}_\text{BS}}}  

\newcommand{\rmv}{\hspace{-0.5mm}}

\begin{document}

\title{RIS Nearfield Position and Velocity Estimation Using a Validated Propagation Model}

\author{\IEEEauthorblockN{
Thomas Zemen\IEEEauthorrefmark{1}, 
Musa Furkan Keskin\IEEEauthorrefmark{2}, 
Moustafa Rahal\IEEEauthorrefmark{3}, 
Thomas Wilding\IEEEauthorrefmark{1}, 
Hamed Radpour\IEEEauthorrefmark{1}, \\ 
Markus Hofer\IEEEauthorrefmark{1}, 
Benoit Denis\IEEEauthorrefmark{4}, 
and Henk Wymeersch\IEEEauthorrefmark{2}}

\IEEEauthorblockA{\IEEEauthorrefmark{1}
Center for Digital Safety \& Security, AIT Austrian Institute of Technology GmbH, Vienna, Austria}
\IEEEauthorblockA{\IEEEauthorrefmark{2}
Department of Electrical Engineering, Chalmers University of Technology, Gothenburg, Sweden}
\IEEEauthorblockA{\IEEEauthorrefmark{3}
Institute for Communication Systems (ICS), University of Surrey, Surrey, UK}
\IEEEauthorblockA{\IEEEauthorrefmark{4}
CEA-Leti, Universit\'e Grenoble Alpes, Grenoble, France}
\IEEEauthorblockA{Email: thomas.zemen@ait.ac.at}
}


\maketitle 

\begin{abstract}
We investigate reconfigurable intelligent surfaces (RISs) for the task of position and velocity estimation in non-LOS (NLOS) indoor scenarios, using a snapshot based multi-step estimation algorithm. We evaluate a compound RIS structure prototype composed of four RIS tiles with \(1\)-bit phase control per RIS unit cell. Numerical simulation results taking the antenna patterns into account are presented for an \(3\,\text{m}\times 3\,\text{m}\) area of interest. We demonstrate that the initial grid search step using the far field assumption is not robust enough for small distances to the RIS center and propose a more robust algorithm. Furthermore, we show that the effect of the antenna pattern causes an increased position and velocity error. Our modified three-step algorithm achieves a position error of \(7\,\)mm and a velocity error of \(0.12\,\)m/s at a distance of \(2\,\)m to the RIS center under a realistic numerical propagation model.
\end{abstract}

\vskip0.5\baselineskip


\section{Introduction} 
Reconfigurable intelligent surfaces (RISs) are used to improve the reliability of millimetre wave (mmWave) wireless communication links in non line-of-sight (NLOS) scenarios by establishing a bypass of the blocking object through an adjustable reflection of the base station (BS) signal to the user equipment (UE). This is especially effective in indoor scenarios such as industrial automation and control. Similarly, a single RIS can also enable the position and velocity estimation of an UE in NLOS of a BS. 

Numerical simulation results and theoretical performance bounds for RIS-based near-field (NF) localization have been presented in \cite{Abu-Shaban21}. A localization method for the NF of a large compound RIS structure in the form of a band has been shown in \cite{Dardari22}. A snapshot based multi-step position and velocity estimation algorithm in NLOS of the BS has been proposed by \cite{Rahal24a}, exploiting the NF properties of the wave field caused by the RIS reflection using a simplified generative model. Corrections for \cite{Rahal24a} have been provided in \cite{Rahal25}. 

The use of RIS in NLOS scenarios for establishing reliable wireless \emph{communication links} between BS and UE has been successfully demonstrated by empirical experiments \cite{Pei21}. In \cite{Radpour24a}, an active RIS for the mmWave frequency band is presented. Numerical simulations and measurements for the received power show a good match for the radiation pattern \cite{Radpour24a} in an automation and control scenario. In \cite{Radpour25} a tracking algorithm for a mobile UE is presented that iteratively updates the reflection coefficient of the RIS unit cells.

In this paper we will utilise a numerical propagation model \cite{Tang22a}, that has been validated by empirical measurements in \cite{Radpour24a, Radpour25}, to explore the downlink RIS-aided NLoS localization use case for indoor scenarios by means of a realistic numerical simulation model. 

\subsection*{Scientific contributions}
\begin{itemize}
\item We analyse the NF three-step localization algorithm from \cite{Rahal25} for a practical compound RIS structure consisting of multiple tiles.

\item We show that the initial grid search (GS) that uses a far-field (FF) assumption for the azimuth and elevation estimation followed by a line search for the range, is not robust for short distances to the RIS center. A robust NF GS step is proposed and analysed.

\item We present position and velocity estimation results for a two-dimensional (2D) area of interest (AOI) using a realistic and empirically validated RIS propagation model \cite{Tang22a, Radpour24a, Radpour25} that includes the antenna patterns of the BS, RIS unit cells and UE. 		
\end{itemize}



\section{System and signal model}
\label{se:SignalModel}
\begin{figure}
	\centering
	\includegraphics[width=0.8\columnwidth]{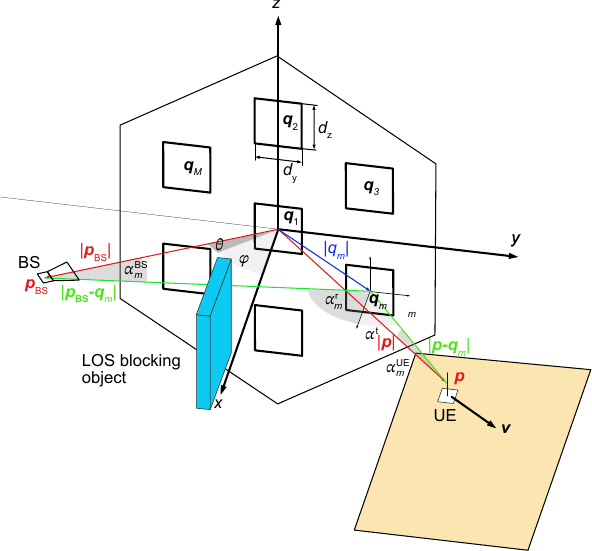}
	\caption{RIS coordinate system for a hexagonal RIS placement in the yz-plane. The BS horn antenna radiates from position $\npBS$ towards the center of the RIS at $\Vec{q}_\text{r}=\Vec{0}=[0,0,0]\oT$ over a distance of $|\npBS|$, similarly the omni-directional UE antenna at position $\Vec{p}$ is within a distance of $|\Vec{p}|$. The LOS link between BS and UE is blocked by an object. The center of the square RIS unit cells are at $\Vec{q}_m$. The RIS is shown enlarged to improve clarity.} 
	\label{fig:RIScoordinates}
\end{figure}

We consider a scenario encompassing a single mobile UE, one BS and one RIS.
The BS is located at $\npBS\in\mathbb{R}^3$ and the UE at $\Vec{p}\in\mathbb{R}^3$ in NLOS with respect to the BS. Both the BS and the UE have direct LOS to a passive reflective RIS with $M$ elements. The center point of a RIS unit cell $m$  is at $\Vec{q}_m\in\mathbb{R}^3$ with $m\in\{1,\ldots, M\}$, see Fig.~\ref{fig:RIScoordinates}. 

The UE is moving linearly with constant velocity $\Vec{v}\in\mathbb{R}^3$ along positions $\Vec{p}_\ell \in\mathbb{R}^3$ given by $\Vec{p}_\ell=\Vec{p}+\Vec{v} \ell \nTS$ where $\ell$ denotes the discrete time index, $\nTS$ the sampling time and $\Vec{p}\in\mathbb{R}^3$ the initial position. The linear motion is assumed for a time interval of length $L \nTS$. Assuming a flat-fading scenario, dominated by the LOS from the BS to the RIS and from the RIS to the UE, the received signal sequence
\be
y_\ell = h_\ell'\sqrt{P_{\text{BS}}} s_\ell + n_\ell\,,
\label{eq:signalModel}
\ee
for $\ell\in\{0,\ldots, L-1\}$, where $L$ denotes the sequence length, $s_\ell$ the transmitted unit-energy downlink (DL) pilot symbols, $P_{\text{BS}}$ the transmit power of the BS, $h_\ell'$ the frequency-flat channel coefficient, and $n_\ell\sim \mathcal{CN}(0,P_n)$ additive complex Gaussian noise, respectively. The noise power $P_n=k_\text{B} T B n_\text{f}$, with the Boltzmann constant $k_\text{B}$, temperature $T$ in Kelvin, bandwidth $B$ and the noise figure $n_\text{f}$.

The channel coefficient $h'_\ell$ is defined as 
\begin{multline}
h'_\ell=\frac{\sqrt{G_{\text{BS}} G_{\text{UE}}}d_\text{y} d_\text{z}}{4\pi}\\
\times \sum_{m=1}^{M}\gamma_{m,\ell}\frac{\sqrt{F_{m}^{\text{c}}(\Vec{p}_\ell)}e^\frac{-j2\pi (\|\npBS-\Vec{q}_m\|+\|\Vec{p}_\ell-\Vec{q}_m\|)}{\lambda}}{\|\npBS-\Vec{q}_m\| \|\Vec{p}_\ell-\Vec{q}_m\|} 
\label{eq:cir}
\end{multline}
where $G_{\text{BS}}$, $G_{\text{UE}}$, $d_\text{y}$, and $d_\text{z}$ are the BS and UE antenna gains and the effective RIS unit cell dimensions in the $y$ and $z$ directions, respectively (see \cite[(9)]{Tang22a}). The complex reflection coefficient of each RIS unit cell $m$ at time index $\ell$ is denoted by $\gamma_{m,\ell}\in\mathcal{A}$, where $\mathcal{A}$ denotes the set of possible phase shifts. We assume randomly chosen and known reflection coefficients $\gamma_{m,\ell} \in \mathcal{A}$ for each RIS unit cell $m$ and time index $\ell$. The wavelength $\lambda=c_0/f$, where $f$ denotes the center frequency, and $c_0$ the speed of light. 

The combined antenna pattern $F_{m}^{\text{c}}(\Vec{p}_\ell)$ of the BS antenna, the RIS unit cell $m$ for receive and transmit operation as well as the UE antenna at position $\Vec{p}_\ell$ is defined as
$$F^\text{c}_m(\Vec{p})=F^\text{BS}(\alpha^\text{BS}_m) F(\alpha^\text{r}_m) F(\alpha^\text{t}_m) F^\text{UE}(\alpha^\text{UE}_m)=$$
$$=\cos(\alpha^\text{BS}_m)^{\frac{G_\text{BS}}{2}-1}\cos(\alpha^\text{r}_m) \cos(\alpha^\text{t}_m) \cos(\alpha^\text{UE}_m)^{\frac{G_\text{UE}}{2}-1} =$$
$$=\left(\frac{\|\npBS\|^2+\|\npBS-\Vec{q}_m\|^2-\|\Vec{q}_m\|^2}{2\|\npBS\|\|\npBS-\Vec{q}_m\|}\right)^{\frac{G_\text{BS}}{2}-1}$$
$$\times \left(\frac{[\npBS]_\text{x}}{\|\npBS-\Vec{q}_m\|}\right)\left(\frac{[\Vec{p}]_\text{x}}{\|\Vec{p}-\Vec{q}_m\|}\right)$$
$$\times
\left(\frac{\|\Vec{p}\|^2+\|\Vec{p}-\Vec{q}_m\|^2-\|\Vec{q}_m\|^2}{2\|\Vec{p}\|\|\Vec{p}-\Vec{q}_m\|}\right)^{\frac{G_\text{UE}}{2}-1} $$
following \cite[(7)]{Tang22a} and the geometric configuration in Fig. \ref{fig:RIScoordinates}. Here, $F^\text{BS}(\cdot)$, $F^\text{UE}(\cdot)$, and $F(\cdot)$ denote the rotation symmetric antenna pattern with respect to the main propagation direction for BS, UE, and RIS unit cell, respectively. The angle with respect to the main wave propagation direction of each antenna is measured by $\alpha^\text{BS}_m$, $\alpha^\text{UE}_m$, $\alpha^\text{r}_m$ and $\alpha^\text{t}_m$ for the BS horn antenna, the UE antenna, the RIS unit cell in reception and the RIS unit cell in transmit direction, respectively.

The numerical model in \eqref{eq:cir} will be used throughout this paper to \emph{generate the channel coefficients} $h_\ell$. We  define the signal to noise ratio (SNR) at the UE as 
$\text{SNR}_\ell = \nicefrac{|h_\ell'|^2 P_\text{BS}}{P_n}$\,.

\section{Near-field position and velocity estimation}
\label{se:PosVelEstimation}
We are interested to utilise the sequence of received signal samples $y_\ell$, $\ell \in \{0,\ldots, L-1\}$ to estimate the unknown position $\Vec{p}$ and velocity $\Vec{v}$ of the UE. 

\subsection{Signal model simplification for position and velocity estimation}
To this end we can simplify \eqref{eq:cir} for the \emph{position and velocity estimation algorithm} with respect to terms that are only affecting the amplitude of $y_\ell$ as follows:
\begin{enumerate}
\item The overall displacement $\Vec{\Delta}_\ell=\Vec{v} \ell \nTS$ is small compared the distance from the UE to the RIS center at $\Vec{q}_\text{r}=[0,0,0]\oT$, i.e.,  $\|\Vec{\Delta}_\ell\| \ll d_\text{r}$ with $d_\text{r}(\Vec{p})=\|\Vec{p}-\Vec{q}_\text{r}\|$.
\item We assume the aperture of the RIS $D=\max \|\Vec{q}_m - \Vec{q}_i\|$ to be much smaller than $d_\text{r}(\Vec{p})$, i.e., $D \ll d_\text{r}(\Vec{p})$.
\item  We assume that the UE's small-scale mobility during the time interval $\Delta t= L\nTS$ mostly induces extra phase changes in the received signal at the UE, while its amplitude remains approximately constant. Under this assumption we can simplify \eqref{eq:cir} by combining the distance terms in the denominator of the sum into a common term in front of the sum (see \eqref{eq:sign_mod_simpl}).
\item We assume constant transmitted symbols $s_\ell=1$ for conciseness, and
\item set the antenna pattern $F_{m}^{\text{c}}(\Vec{p}_\ell)=1$.
\end{enumerate}

With these assumptions we can simplify \eqref{eq:cir} as follows; 
\begin{multline}
y_\ell \approx \underbrace{\frac{\sqrt{P_{\text{BS}} G_{\text{BS}} G_{\text{UE}}}d_\text{y} d_\text{z} }{4\pi \|\npBS-\Vec{q}_\text{r}\|\|\Vec{p}-\Vec{q}_\text{r}\|} e^\frac{-j2\pi d_\text{r}(\Vec{p})}{\lambda} }_{\alpha} \\
\times\rmv\rmv \sum_{m=1}^{M}\underbrace{\gamma_{m,\ell}e^\frac{-j2\pi \|\npBS-\Vec{q}_m\|}{\lambda}}_{w_{m,\ell}} \underbrace{e^{\frac{-j2\pi}{\lambda}\left[\|\Vec{p}+\Vec{v}\ell\nTS-\Vec{q}_m\|- d_\text{r}(\Vec{p})\right]} }_{a_{m,\ell}(\Vec{p},\Vec{v})}+ n_\ell \\
= \alpha \underbrace{\Vec{w}_\ell\oT \Vec{a}_\ell(\Vec{p},\Vec{v})}_{h_\ell(\Vec{p},\Vec{v})} +n_\ell = \alpha h_\ell(\Vec{p},\Vec{v})+ n_\ell 
\label{eq:sign_mod_simpl} 
\end{multline} 
where $\Vec{w}_\ell=[w_{1,\ell}, \ldots, w_{M,\ell}]\oT\in\mathbb{C}^M$ and $\Vec{a}_\ell= [a_{1,\ell}(\Vec{p},\Vec{v}), \ldots, a_{M,\ell}(\Vec{p},\Vec{v})]\oT\in\mathbb{C}^M$. 

The position and velocity estimation algorithm from \cite{Rahal25}, employs several additional approximation steps including the small angle approximation (SAA) of the exponential term, $e^{j x} \approx 1+ j x$ for $x\ll 1$. Hence, on the right hand side of \eqref{eq:sign_mod_simpl} we subtract the common propagation distance $d_\text{r}(\Vec{p})$ from the center of the RIS to the position of the UE in the exponential term, while a common phase term $e^{\nicefrac{-j2\pi d_\text{r}(\Vec{p})}{\lambda}}$ is added before the sum to keep the propagation equation physically correct. 

Combining $L$ time samples we can vectorise \eqref{eq:sign_mod_simpl} as
\be
\Vec{y}= \alpha \Vec{h}(\Vec{p},\Vec{v})+ \Vec{n} \in\mathbb{C}^L\,.
\label{eq:SignalModel}
\ee
Conditional on the knowledge of $\Vec{p}$ and $\Vec{v}$, we can obtain a conditional maximum likelihood (ML) estimate of $\alpha$ as
\be
\hat{\alpha} = \frac{\Vec{h}(\Vec{p},\Vec{v})\oH}{\|\Vec{h}(\Vec{p},\Vec{v})\|^2} \Vec{y}\,. 
\label{eq:alpha}
\ee
Finally, by inserting \eqref{eq:alpha} into \eqref{eq:SignalModel} the ML estimator can be expressed as
\begin{align}
(\hat{\Vec{p}}, \hat{\Vec{v}}) &= \argmin_{\Vec{p}, \Vec{v}} \left\|\left(\Vec{I} - \frac{\Vec{h}(\Vec{p},\Vec{v}) \Vec{h}(\Vec{p},\Vec{v})\oH}{\|\Vec{h}(\Vec{p},\Vec{v})\|^2}\right) \Vec{y}\right\|^2 \nonumber\\
&= \argmin_{\Vec{p}, \Vec{v}} \left\|{\Vec{\Pi}^\bot_{\Vec{h}(\Vec{p},\Vec{v})}}\Vec{y}\right\|^2\,,
\label{eq:ML}
\end{align}
where $\Vec{\Pi}^\bot_{\Vec{x}}= \Vec{I}- \frac{\Vec{x}\Vec{x}\oH}{\|\Vec{x}\|^2}$ denotes the projection on the null space of $\Vec{x}$.

\subsection{Reduced-complexity algorithm}
To reduce the complexity of the 6D ML search required in \eqref{eq:ML}, the authors of \cite{Rahal25} propose an algorithm composed of three-steps, which are summarised below with our modifications.

\subsubsection{Initial position estimation using a grid search}
For the initial GS we represent the position vector $\Vec{p}(\varphi, \theta, r)$ of the UE in spherical coordinates with $\varphi$, $\theta$ and $r$ denoting azimuth, elevation and range.

In \cite[Alg. 2]{Rahal25} an initial 2D GS with respect to azimuth and elevation, using the FF approximation of the RIS phase response $\Vec{a}_\ell(\Vec{p},\Vec{v)}$ and assuming $\Vec{v}_0=\Vec{0}$ is described. This step is followed by a line search to obtain the range estimate. 

For the RIS structure used in our experimental setup, this approach leads to a large error for the range estimate. Hence, we replace the initial position estimation step by a 3D GS in azimuth, elevation and range using the NF expression in \eqref{eq:sign_mod_simpl}. A coarse grid can be used initially, due to the subsequent refinement steps explained below. Overall, our modification provides a smaller initial position error compared to the first algorithm step in \cite{Rahal25}, while the complexity increases only moderately. Assuming $\Vec{v}^{(0)}=\Vec{0}$ we perform a coarse GS using,
$(\hat{\varphi}, \hat{\theta}, \hat{r}) = \argmin_{\varphi, \theta, r} \left\|{\Vec{\Pi}^\bot_{\Vec{h}(\Vec{p}(\varphi, \theta, r),\Vec{v}^{(0)}}}\Vec{y}\right\|^2$\,,
providing an initial position estimate ${\Vec{p}}^{(0)}=\Vec{p}(\hat{\varphi}, \hat{\theta}, \hat{r})$. 

\subsubsection{Closed-form position and velocity refinement}
In \cite{Rahal25}, a linearisation of the RIS phase response $a_{m,\ell}(\Vec{p},\Vec{v})$ is derived assuming an available position estimate $\Vec{p}^{(i)}$ and velocity estimate $\Vec{v}^{(i)}$. In \cite[Alg. 3]{Rahal25} the initial position estimate $\Vec{p}^{(i)}$ is refined by adding a position displacement $\Vec{p}_{\delta}(\Vec{y}, \hat{\Vec{p}}^{(i)}, \hat{\Vec{v}}^{(i)})$ obtained from a closed-form (CF) expression taking the received signal vector $\Vec{y}$ as input in addition to the current position $\hat{\Vec{p}}^{(i)}$ and velocity estimate $\hat{\Vec{v}}^{(i)}$ at iteration $(i)$, which yields the updated position estimate as
\be
\hat{\Vec{p}}^{(i+1)} = \hat{\Vec{p}}^{(i)} + \Vec{p}_{\delta}(\Vec{y}, \hat{\Vec{p}}^{(i)}, \hat{\Vec{v}}^{(i)})\,.
\label{eq:PosRef}
\ee

Similarly in \cite[Alg. 4]{Rahal25} the initial velocity estimate is refined by adding a velocity offset $\Vec{v}_\delta(\Vec{y}, \hat{\Vec{p}}^{(i)}, \hat{\Vec{v}}^{(i)})$ obtained from a CF expression leading to
\be
\hat{\Vec{v}}^{(i+1)} = \hat{\Vec{v}}^{(i)} + \Vec{v}_\delta(\Vec{y}, \hat{\Vec{p}}^{(i)}, \hat{\Vec{v}}^{(i)})\,.
\label{eq:VelRef}
\ee
The position and velocity refinements are iterated until \eqref{eq:ML} converges at iteration $i'$.

\subsubsection{6D gradient descent search}
Finally, a 6D gradient descent search starting from $(\hat{\Vec{p}}^{(i')}, \hat{\Vec{v}}^{(i')})$ is performed to solve \eqref{eq:ML} and obtain $(\hat{\Vec{p}}, \hat{\Vec{v}})$ as described in \cite{Rahal25}.

\section{Numerical simulation setup and results}
\label{se:Results}

\subsection{Parameter configuration and implementation details}
\begin{figure}
	\centering
	\includegraphics[width=0.8\columnwidth]{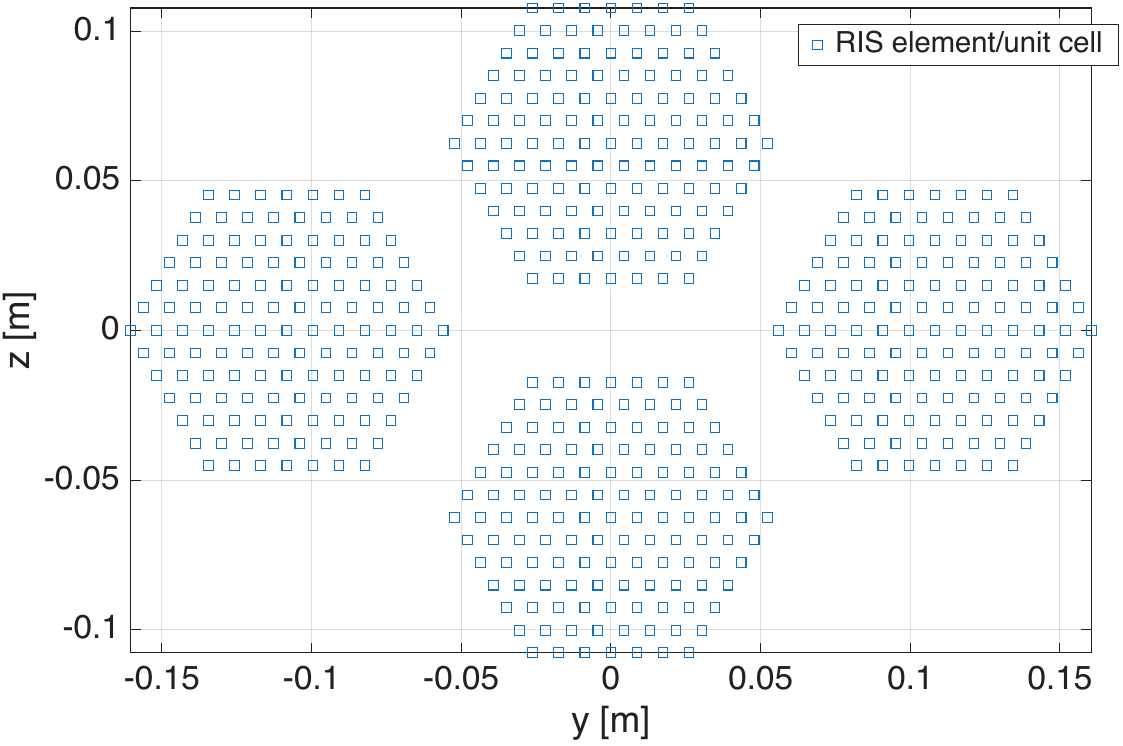}
	\caption{RIS composed of four RIS tiles with each 127 elements, resulting in a total number of $M=508$ elements.} 
	\vspace{-0.3cm}
	\label{fig:RIStiles}
\end{figure}

In \cite{Radpour24a, Radpour25} a mmWave RIS testbed is described that consists of a RIS with $M=127$ elements. The position and velocity estimation algorithm in Sec. \ref{se:PosVelEstimation} exploits the NF properties that are observable for distances below the Fraunhofer distance defined by $d_\text{F}=\nicefrac{2D^2}{\lambda}$ where $D$ is the aperture of the RIS. To increase the aperture $D$, we combine four RIS tiles with 127 elements each, into a larger composite RIS with a total number of $M=508$ RIS unit cells, see Fig.~\ref{fig:RIStiles}.
The composite RIS for the center frequency $f=23.8\,$GHz has a Fraunhofer distance $d_\text{F} = 16.3$\,m, which is sufficient for indoor position and velocity estimation applications in industrial automation and control scenarios. Each RIS unit cell can be controlled by one bit to have a reflection coefficient $\gamma\in\mathcal{A}=\{\me^{j 2 \pi\frac{-15^\circ}{360^\circ}}, \me^{j 2 \pi \frac{165^\circ}{360^\circ}}\}$. The detailed parameters can be found in Table~\ref{tab:systemparam}.
\begin{table}
\begin{center}
\caption{RIS parameters and simulation setup.}
\label{tab:systemparam}
\begin{tabular}{ll} 
\toprule
Parameter	      	&  Definition\\
\midrule
$f = 23.8$\,GHz		& 	 center frequency \\
$M = 4\times 127=508$    & number of RIS unit cells\\
$d_z = d_y = 6.6$\,mm	& 	 effective RIS unit cell size \\
$d = 8.7\,\text{mm} \approx 0.7 \lambda$ & smallest RIS unit cell distance\\
$\mathcal{A}=\{\me^{\mj 2 \pi\frac{-15^\circ}{360^\circ}}, \me^{\mj 2 \pi \frac{165^\circ}{360^\circ}}\}$ & RIS reflection coefficient set\\
\midrule
$P_\text{BS}=20$\,dBm		& 	 BS transmit power\\
$G_\text{BS} = 19$\,dB		&  BS antenna gain\\
$G_\text{UE} = 3.2$\,dB		&   UE monopole antenna gain\\
$n_\text{f} = 8$\,dB	    & noise figure\\
$T= 293$\,K & temperature\\	
$B = 1$\,MHz & bandwidth\\
\midrule
$\Vec{p}_\text{BS}=[1,-3,3]\oT$\,m & BS position\\
$\Vec{v}=[1,-1,2]\oT/\sqrt{6}$\,m & UE velocity\\
$\Vec{q}_\text{r}=[0,0,0]\oT$\,m & reference position in center of RIS\\
$\nTS = 0.1$\,ms & sampling time\\ 
$L = 40$ & number of samples\\
$F=250$ & number of simulation runs\\
\bottomrule
\end{tabular}
\end{center}
\vspace{-0.3cm}
\end{table}

\begin{figure}
	\centering
	\includegraphics[width=0.8\columnwidth]{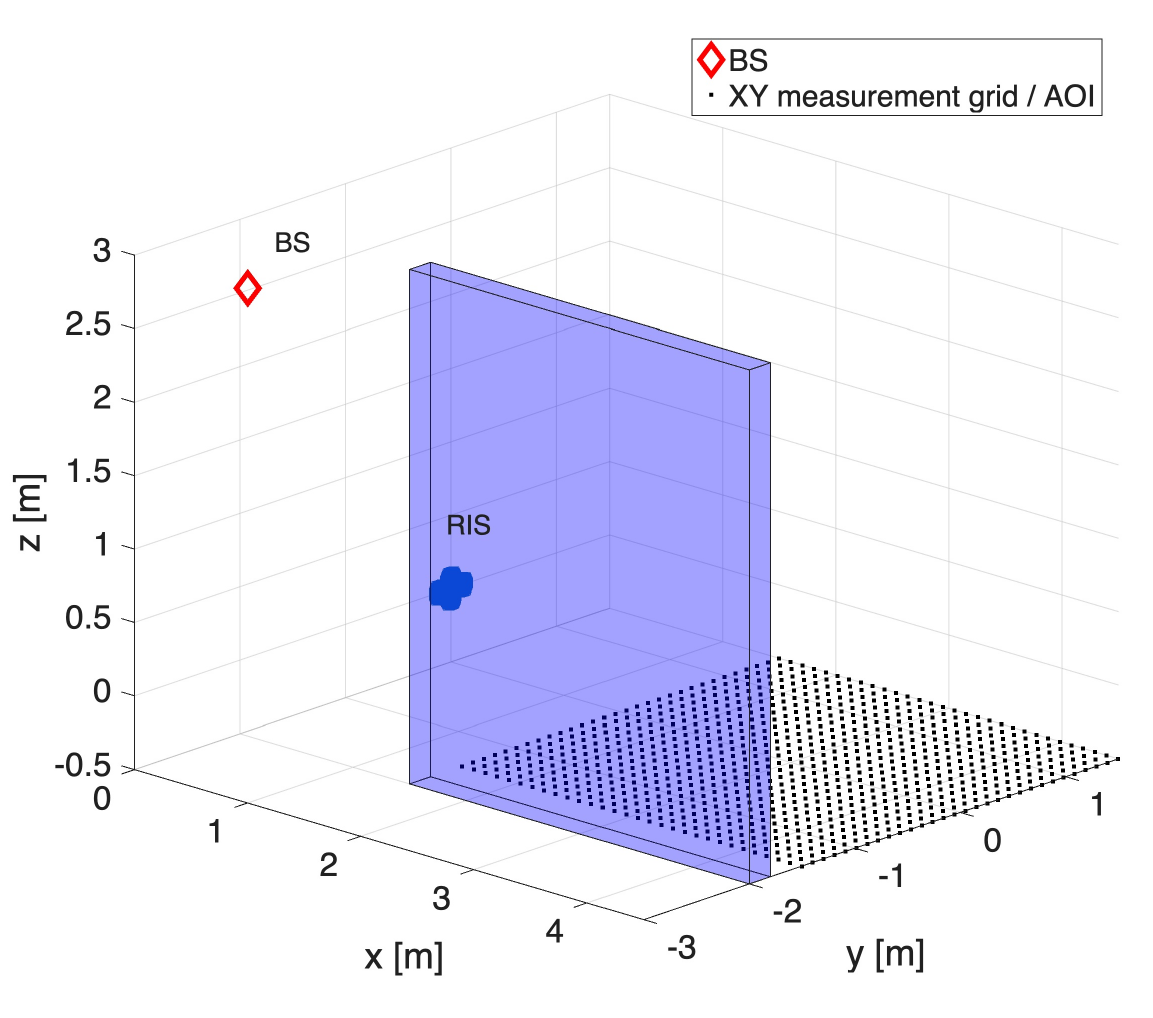}
	\caption{Geometrical configuration of RIS, BS and AOI for the UE. The translucent blue cube blocks the LOS between BS and UE.} 
	\vspace{-0.3cm}
	\label{fig:RISexperiment}
\end{figure}

The geometrical configuration of the RIS experiment is depicted in  Fig.~\ref{fig:RISexperiment}.
The numerical simulation evaluates a 2D area of interest (AOI) with a size of $3\,\text{m} \times 3\,$m using a grid spacing of $10\,$cm in $x$ and $y$ direction at height $z=-0.5$\,m. The grid points serve as initial position $\Vec{p}$ and the UE velocity is set to $\Vec{v}=[1,-1,2]\oT/\sqrt{6}$\,m. 

For the initial coarse GS we define the search space as a 3D discrete set defined by $\phi\in\{-70:1:70\}\,^\circ$, $\theta\in\{-70:1:70\}\,^\circ$, and $r\in\{0.1:0.2:4\}\,\text{m}$ using Matlab notation. The limited range in azimuth and elevations is chosen since the RIS prototype, that we model in the simulation environment, uses an element spacing of $0.7 \lambda > \lambda/2$.  To further improve the coarse GS position estimate we perform a finer GS step using a 3D set defined as $\phi\in\{\hat{\phi}-1:0.1:\hat{\phi}+1\}^\circ$, $\theta\in\{\hat{\theta}-1:0.1:\hat{\theta}+1\}^\circ$, and $r\in\{\hat{r}-0.2:0.005:\hat{r}+0.2\}\,\text{m}$.

We perform Monte Carlo simulations with $F=250$ runs to obtain the root mean square (RMS) position error (RMSPE) and RMS velocity error (RMSVE). The reflection coefficients $\gamma_{m,\ell}$, $m\in\{1,\ldots,M\}$, $\ell\in\{0,\ldots,L-1\}$ are randomly chosen for each run. 

\subsection{Simulation results and dicussion}
\begin{figure}
	\centering
	\includegraphics[width=0.7\columnwidth]{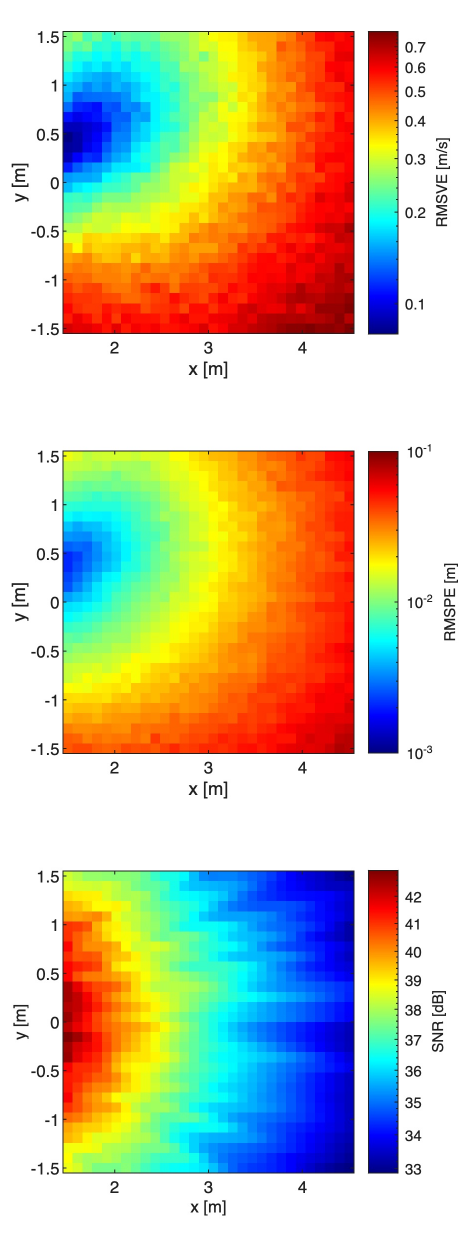}
	\caption{RMSVE, RMSPE, and average SNR in the 2D AOI. The RMS errors are obtained after the 6D gradient descent search averaged over $F=250$ runs.}
	\vspace{-0.5cm} 
	\label{fig:RMSEAOI}
\end{figure}

The RMSVE and the RMSPE as well as the average SNR for the 2D AOI are shown in Fig. \ref{fig:RMSEAOI}. Clearly the RMSVE and the RMSPE increase with distance from the RIS.\footnote{The obtained RMSVE is smaller than the simulation results shown in \cite{Rahal25} although the number of RIS unit cells is smaller in our setup. This is due to the mistake in \cite[(4)]{Rahal24a}, that was corrected in \cite{Rahal25} but the simulation results have not been updated in \cite{Rahal25}.} This can be attributed to two effects: (i) the NF effect, i.e., the wavefront curvature, decays and (ii) the signal strength decreases with distance.  
We want to point out, that the RMSVE and RMSPE show similar distribution but differ from the SNR distribution in the 2D AOI. Hence, the geometric properties of the measurement setup have a strong influence on the RMSVE and RMSPE, see also \cite[App. E]{Rahal25}. 

\begin{figure}
	\centering
	\includegraphics[width=0.8\columnwidth]{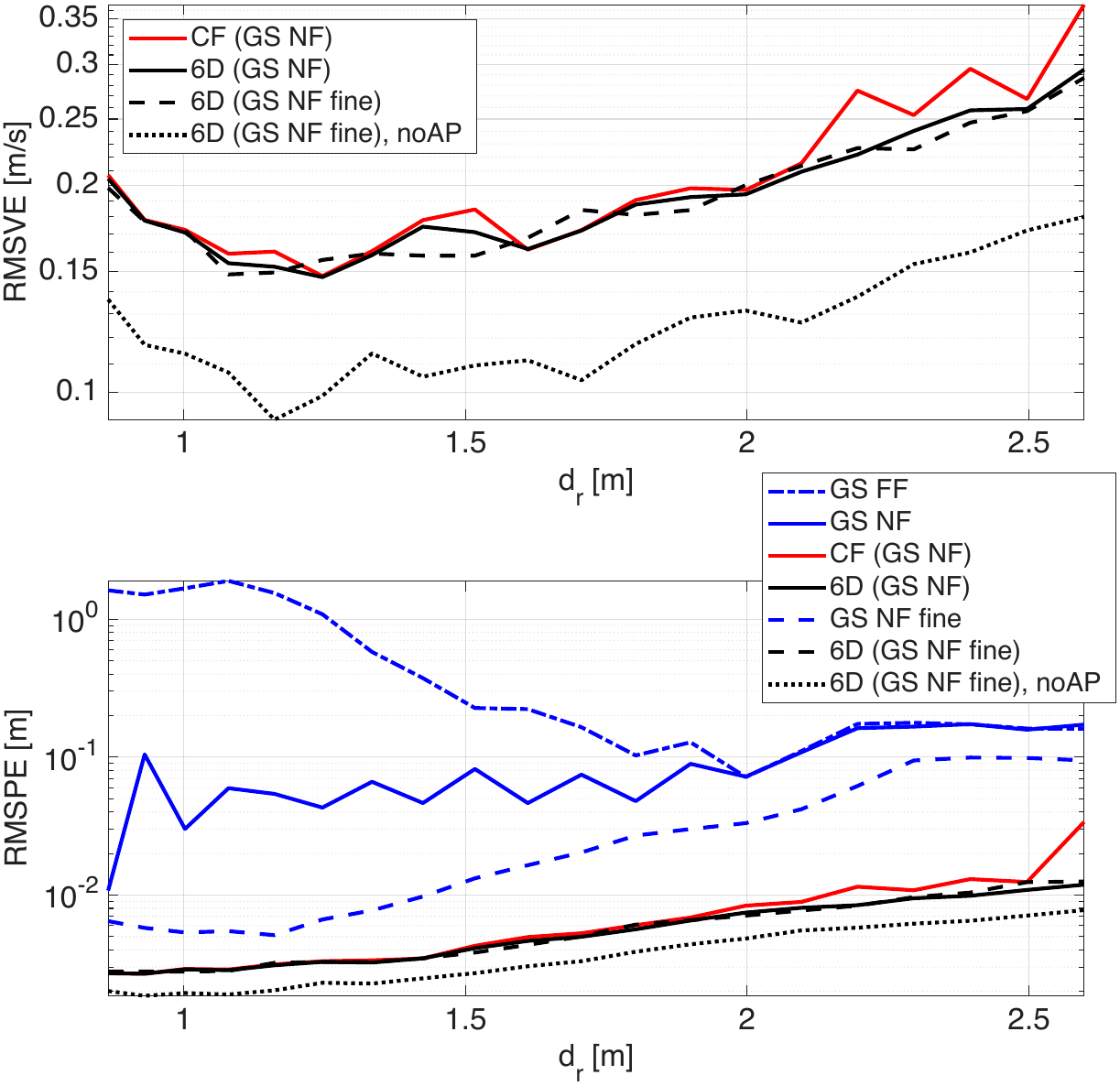}
  \caption{RMSVe and RMSPE versus the distance to the center of the RIS. We evaluate the three algorithm steps (i) grid search (GS), (ii) closed-form refinement (CF), and (iii) 6D gradient decent search (6D). We compare the GS using the FF  assumption (GS FF) with the more robust GS using the NF assumption (GS NF). Furthermore we show the GS in the NF  followed by a finer GS step (GS NF fine). Finally, we also depict the result without antenna pattern (6D (GS NF fine), noAP).}
  \vspace{-0.5cm}
	\label{fig:RMSEvsx}
\end{figure}

In Fig. \ref{fig:RMSEvsx}, we present the RMS errors after the individual steps of the estimation algorithm (i) GS, (ii) CF refinement, and (iii) 6D gradient descent. Here we plot the RMS error versus $d_r(\Vec{p})$. We evaluate a line parallel to the $x$-axis in the middle of the AOI with $y=0$\,m and $z=-0.5$\,m

First, we analyse the initial GS step using \cite[Alg. 2]{Rahal25} where a GS using the FF assumption (GS FF) estimates azimuth $\phi$ and elevation $\theta$ followed by a line search to estimated the range $r$. The numerical result in Fig. \ref{fig:RMSEvsx} show that this approach leads to error propagation for shorter distances where the FF assumption is clearly violated. Hence, the proposed robust NF GS (GS NF) shows a smaller error for distances below $2$\,m. A following refinement step with a finer grid can further reduce the RMSPE, see label `GS NF fine' in Fig. \ref{fig:RMSEvsx}.

Second, the CF refinement step provides a further reduction of the position and velocity error. Here the clear dependence on the distance to the RIS is visible, see label `CF (GS NF)'. 

Third, the 6D gradient descent search provides the final estimate, achieving a position error of $7\,$mm and a velocity error of $0.12\,$m/s in a distance of $2\,$m to the RIS. It turns out that CF and 6D steps lead to the same RMSVE and RMSPE for both initial GS algorithm steps, `GS NF' and `GS NF fine'. Hence, the additional effort for the fine GS step is not necessary and can be omitted.

Finally we depict the RMSVE and RMSPE for $F_{m}^{\text{c}}(\Vec{p}_\ell)=1$ in \eqref{eq:cir}, i.e. no antenna pattern (noAP) is considered. The result shows, that the antenna pattern increases the estimation error, since it is not included in the position and velocity estimation algorithm and causes an attenuation of the received signal. However, this small increase does not hinder the application for practical use cases such as industrial automation and control.
 
In Fig. \ref{fig:RMSEvsx} we can see that the RMSVE increases for distances shorter than $1.2\,$m and similarly the RMSPE flattens for this distance range. This is due to the closed-form refinement expressions derived in \cite{Rahal25} for \eqref{eq:PosRef} and \eqref{eq:VelRef}, which allows the application of the proposed algorithm for $d_\text{r}(\Vec{p})\gg \|\Delta_\ell\|$.

\section{Conclusion} 
\label{se:Conclusion}
We analysed and improved a position and velocity estimation algorithm for a RIS in NLOS that uses a single snapshot sequence of DL pilot symbols. The multi-step algorithm from \cite{Rahal25} is numerically evaluated using a realistic propagation model that incorporates the antenna pattern and a practical placement of RIS unit cells. A compound RIS consisting of four RIS tiles is used with a Fraunhofer distance of $16$\,m. We modify the initial algorithm step in \cite{Rahal25} to perform a 3D NF GS that is followed by a CF refinement and finally a 6D gradient descent search, achieving a position error of $7\,$mm and a velocity error of $0.12\,$m/s in a distance of $2\,$m to the RIS. A detailed analysis of the modified algorithm steps is provided as well as a 2D RMSVE and RMSPE evaluation.

\section*{Acknowledgment}
The work of T. Zemen, T. Wilding, H. Radpour, and M. Hofer is funded within the Principal Scientist grant Dependable Wireless 6G Communication Systems (DEDICATE 6G).



\end{document}